\documentclass{KapEdbk} % Computer Modern font calls

\kluwerbib
\usepackage{graphicx}
\begin{document}
\articletitle{A brief history of drop formation}

\author{Jens Eggers}
\affil{
School of Mathematics, 
University of Bristol, University Walk, \\
Bristol BS8 1TW, United Kingdom 
     }
\email{jens.eggers@bris.ac.uk}

\begin{abstract}
Surface-tension-related phenomena have fascinated researchers 
for a long time, and the mathematical description pioneered 
by Young and Laplace opened the door to their systematic study. 
The time scale on which surface-tension-driven motion takes 
place is usually quite short, making experimental investigation
quite demanding. Accordingly, most theoretical and experimental
work has focused on static phenomena, and in particular the 
measurement of surface tension, by physicists like E\"otv\"os, 
Lenard, and Bohr. Here we will review some of the work that has
eventually lead to a closer scrutiny of time-dependent flows,
highly non-linear in nature. Often this motion is 
self-similar in nature, such that it can in fact be mapped onto
a pseudo-stationary problem, amenable to mathematical analysis.
\end{abstract}


\section*{Introduction}
Flows involving free surfaces lend themselves to observation, 
and thus have been scrutinized for hundreds of years.
The earliest theoretical work was concerned 
almost exclusively with the equilibrium shapes of fluid bodies, 
and with the stability of the motion around those shapes. 
Experimentalists, always being confronted with physical reality, 
were much less able to ignore the strongly non-linear nature 
of hydrodynamics. Thus many of the non-linear phenomena, that are
the focus of attention today, had already been reported 170 years ago. 
However, with no theory in place to put these observations into 
perspective, non-linear phenomena took the back seat to other issues,
and were soon forgotten. Here we report on the 
periodic rediscovery of certain non-linear features of drop formation,
by retracing some of the history of experimental observation of 
surface tension driven flow. Recently there has been some progress 
on the theoretical side, which relies on the self-similar nature 
of the dynamics close to pinching. 

\section{Savart and Plateau}
Modern research on drop formation begins with the seminal 
contribution of \cite{S1833}. He was the first to 
recognize that the breakup of liquid jets is governed by
laws independent of the circumstance under which the jet
is produced, and concentrated on the simplest possible case
of a circular jet. Without photography at one's disposal,
experimental observation of drop breakup is very difficult,
since the timescale on which it is taking place is very short. 

\begin{figure}
\begin{center}
\includegraphics[width=0.09\hsize,angle=90]{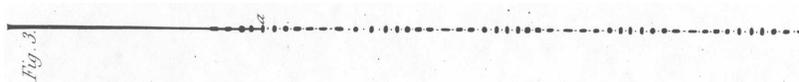}
\end{center}

\caption{\label{fig1} 
A figure from Savart's original paper (\cite{S1833}), showing the
breakup of a liquid jet 6 mm in diameter. It clearly shows the 
succession of main and satellite drops as well as drop oscillations.
  }
\end{figure}

Yet Savart was able to extract a remarkably accurate and complete
picture of the actual breakup process using his naked eye alone.
To this end he used a black belt, interrupted by narrow white
stripes, which moved in a direction parallel to the jet. This 
effectively allowed a stroboscopic observation of the jet. 
To confirm beyond doubt the fact that the jet breaks up into
drops and thus becomes discontinuous, Savart moved a 
``slender object'' swiftly across the jet, and found that 
it stayed dry most of the time. Being an experienced swordsman,
he undoubtedly used this weapon for his purpose (\cite{Cc}).
Savart's insight into the {\it dynamics} of breakup is best
summarized by Fig.\ref{fig1} taken from his paper (\cite{S1833}). 

To the left one sees the continuous jet as it leaves the nozzle.
Perturbations grow on the jet, until it breaks up into drops, at 
a point labeled ``a''. Near a an elongated neck has formed between 
two bulges which later become drops. After breakup, in between two 
such drops, a much smaller ``satellite'' drop is always visible. 
Owing to perturbations received when they were formed, the drops 
continue to oscillate around a spherical shape. Only the very last
moments leading to drop formation are not quite resolved in 
Fig.\ref{fig1}. 

\begin{figure}
\begin{center}
\includegraphics[width=0.6\hsize,angle=-90]{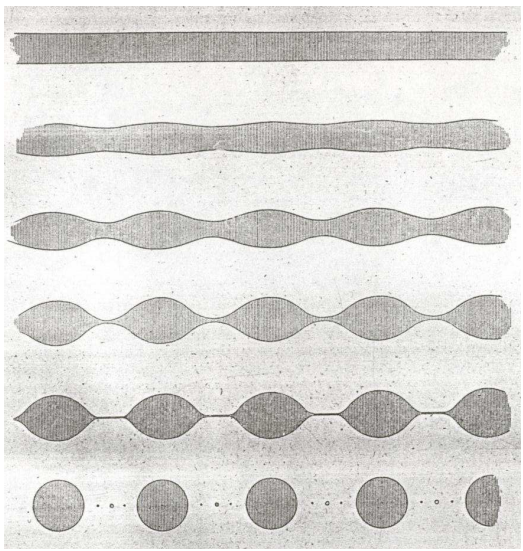}
\end{center}
\caption{\label{Plateau} 
Breakup of a liquid column of oil, suspended in a mixture of 
alcohol and water (\cite{P1849}). First small perturbations grow,
leading to the formation of fine threads. The threads each break
up leaving three satellites.
  }
\end{figure}

From a theoretical point of view, what is missing is the realization
that surface tension is the driving force behind drop breakup, the
groundwork for the description of which was laid by \cite{Y1804} and
\cite{L1805}. Savart however makes reference to mutual attraction
between molecules, which make a sphere the preferred shape, around
which oscillations take place. The crucial role of surface tension 
was recognized by \cite{P1849}, who confined himself mostly 
to the study of equilibrium shapes. This allows one to predict 
whether a given perturbation imposed on a fluid cylinder will 
grow or not. Namely, any perturbation that will lead to a reduction 
of surface area is favored by surface tension, and will thus grow.
This makes all sinusoidal perturbations with wavelength longer
than $2\pi$ unstable. At the same time as Plateau, Hagen published
very similar investigations, without quite mastering the mathematics
behind them (\cite{H1849}). The ensuing quarrel between the two 
authors, published as letters to {\it Annalen der Physik}, is quite
reminiscent of similar debates over priority today. 

A little earlier Plateau had developed his own experimental technique
to study drop breakup (\cite{P1843}), by suspending a liquid bridge in 
another liquid of the same density in a so-called ``Plateau tank'', 
thus eliminating the effects of gravity. Yet this research was focused
on predicting whether a particular configuration would be stable 
or not. However Plateau also included some experimental sketches 
(cf. Fig.\ref{Plateau}) that offer interesting insight into the 
nonlinear dynamics of breakup for a viscous fluid:
first a very thin and elongated thread forms, which has its minimum in the 
middle. However, the observed final state of a satellite drop in the
center, with even smaller satellite drops to the right and left 
indicates that the final stages of 
breakup are more complicated: the thread apparently broke at 4 different
places, instead of in the middle. 

Following up on Plateau's insight, \cite{R1879} added
the flow dynamics to the description of the breakup process. At low
viscosities, the time scale $\tau$ of the motion is set by a balance of 
inertia and surface tension:
\begin{equation}
\label{tau}
\tau=\sqrt{\frac{r^3\rho}{\gamma}}. 
\end{equation}
Here $r$ is the radius of the (water) jet, $\rho$ the density, and 
$\gamma$ the surface tension. For the jet shown in Fig.\ref{fig1}, this
amounts to $\tau=0.02\; s$, a time scale quite difficult to observe 
with the naked eye. 
Rayleigh's linear stability calculation of a fluid cylinder only allows
to describe the initial growth of instabilities as they initiate
near the nozzle. It certainly fails to describe the details of drop
breakup leading to, among others, the formation of satellite drops. 
Linear stability analysis is however quite a good predictor of 
important quantities like the continuous length of the jet.

\begin{figure}
\begin{center}
\includegraphics[width=0.8\hsize]{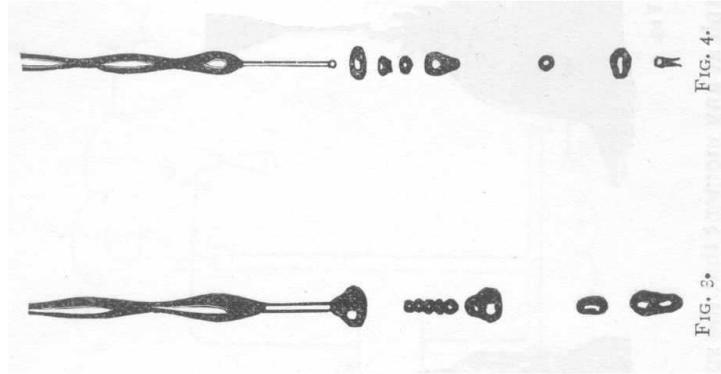}
\end{center}
\caption{\label{Rayleigh} 
Two photographs of water jets taken by \cite{R1891}, using 
a short-duration electrical spark. 
  }
\end{figure}

\begin{figure}
\begin{center}
\includegraphics[width=1\hsize]{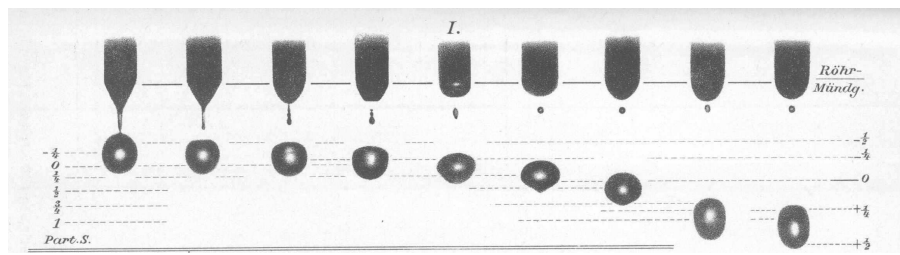}
\end{center}
\caption{\label{lenard} 
A sequence of pictures of a drop of water falling from a pipette
(\cite{L1887}). For the first time, the sequence of events leading
to satellite formation can be appreciated. 
  }
\end{figure}

\section{Photography}

Rayleigh was well aware of the intricacies of the last stages
of breakup, and published some experimental pictures himself (\cite{R1891}).
Unfortunately, these pictures were produced by a single short spark,
so they only transmit a rough idea of the {\it dynamics} of the 
process. However, it is again clear that satellite drops, or entire 
sequences of them, are produced by elongated necks between two main
drops. Clearly, what is needed for a more complete understanding is 
a sequence of photographs showing one to evolve into the other. 

The second half of the 19th century is an era that saw a great 
resurgence of the interest in surface tension related phenomena,
both from a theoretical and experimental point of view. The driving
force was the central role it plays in the quest to understand the 
cohesive force between fluid particles (\cite{R02}), for example by 
making precise measurements of the surface tension of a liquid. 
Many of the most well-known physicists of the day contributed to
this research effort, some of whom are known today for their later 
contributions to other fields (\cite{E1886,Q1877,L1887,B09}). 
A particular example is the paper by
\cite{L1887}, who observed the drop oscillations that remain after
break-up, already noted by Savart. By measuring their 
frequency, the value of the surface tension can be deduced. 

To record the drop oscillations, Lenard used a stroboscopic 
method, which allows to take an entire sequence with a time resolution
that would otherwise be impossible to achieve.
As more of an aside, Lenard also records a sequence showing
the dynamics close to breakup, leading to the separation of a drop.
It shows for the first time the origin of the satellite drop:
first the neck breaks close to the main drop, but before it is able
to snap back, it also pinches on the side toward the nozzle. 
The presence of a slender neck is intimately linked to the profile 
near the pinch point being very asymmetric: on one side it is very 
steep, fitting well to the shape of the drop. On the other side it
is very flat, forcing the neck to be flat and elongated. 

\begin{figure}
\begin{center}
\includegraphics[width=5cm]{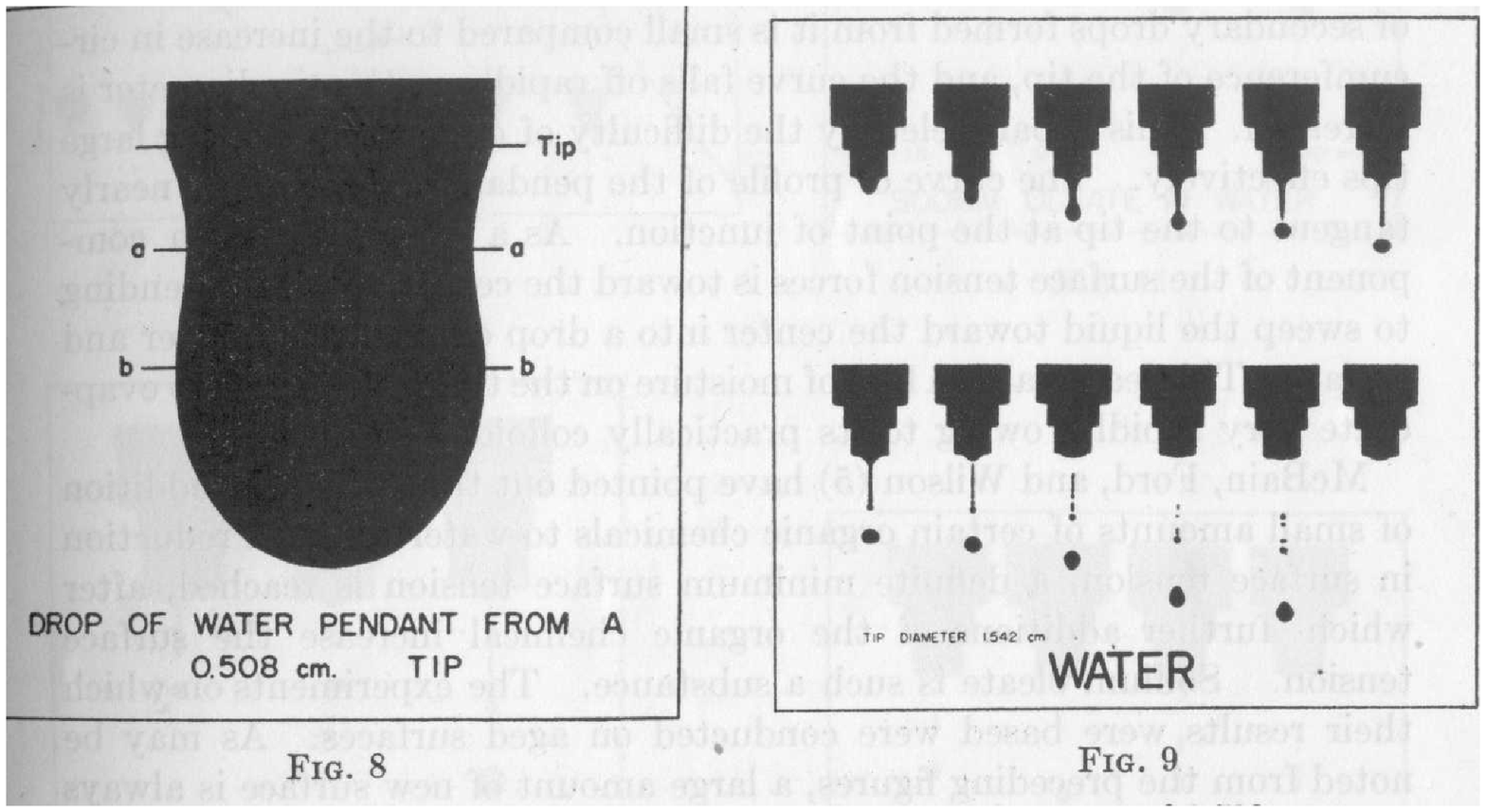}
\hspace{1cm}
\includegraphics[width=5cm]{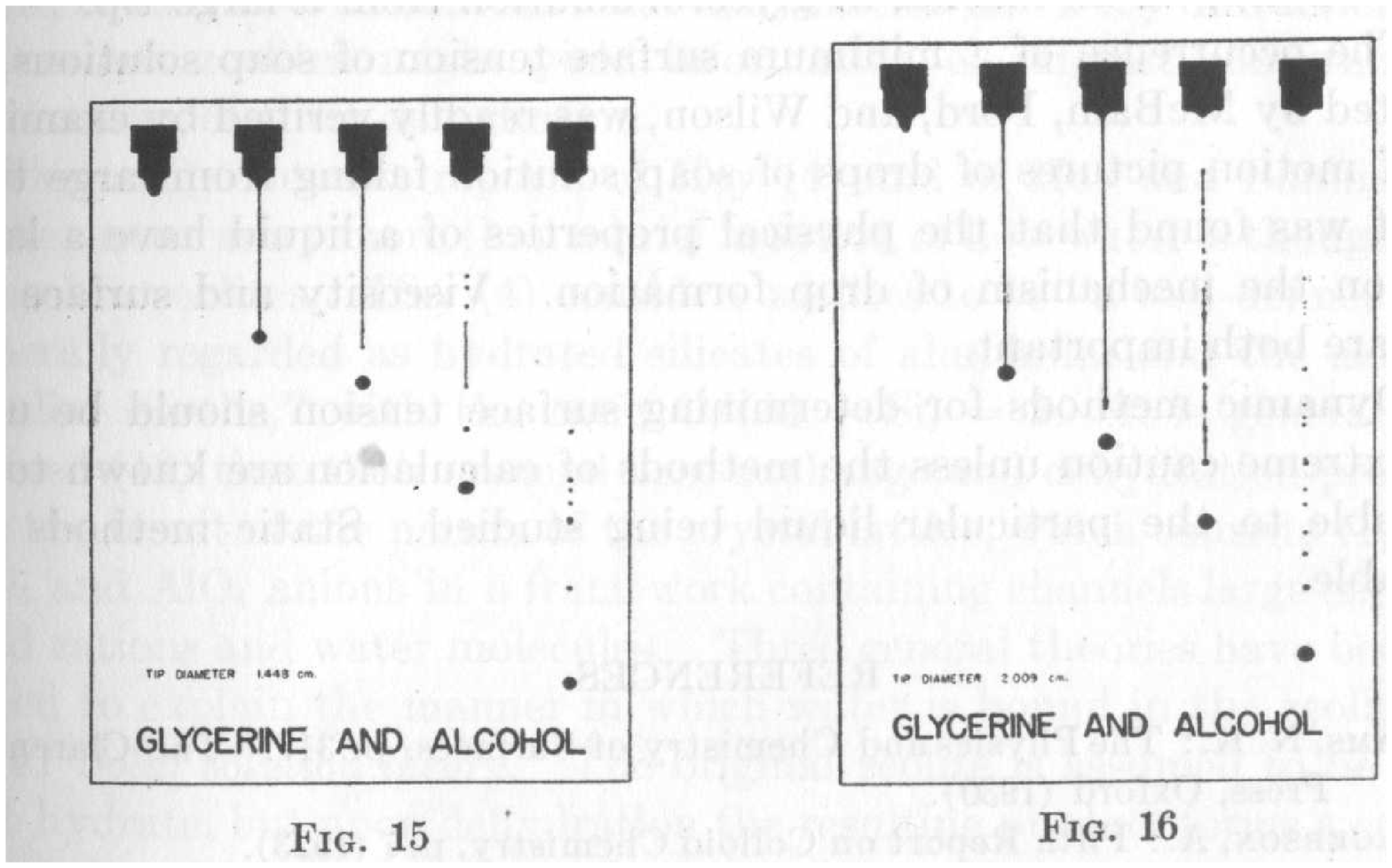}
\end{center}
\caption{\label{edgerton} 
A drop of water (left) and a glycerol-alcohol mixture (right)
falling from a pipette (\cite{EHT37}). The drop of viscous fluid pulls 
out long necks as it falls.
  }
\end{figure}

However, as noted before, few people took note of the fascinating
dynamics close to breakup. From a theoretical point of
view, tools were limited to Rayleigh's linear stability analyses, 
which does not allow to understand satellite formation. Many years later,
the preoccupation was still to find simple methods to measure surface 
tension, one of them being the ``drop weight method'' (\cite{HB19}).
The idea of the method is to measure surface tension by measuring 
the weight of a drop falling from a capillary tubes of defined diameter.
Harold Edgerton and his colleagues looked at time sequences of a drop
of fluid of different viscosities falling from a faucet (\cite{EHT37}),
rediscovering some of the features observed originally by Lenard, but adding
some new insight. 

Fig.\ref{edgerton} shows a water drop falling from a faucet, forming
quite an elongated neck, which then decays into {\it several} satellite
drops. The measured quantity of water thus comes from the main drop as
well as from some of the satellite drops; some of the satellite drops are
projected upward, and thus do not contribute. The total weight thus 
depends on a very subtle dynamical balance, that can hardly be a reliable 
measure of surface tension. In addition, as Fig.\ref{edgerton} demonstrates,
a high viscosity fluid like glycerol forms extremely long threads, 
that break up into myriads of satellite drops. In particular, the drop
weight cannot be a function of surface tension alone, but also depends
on viscosity, making the furnishing of appropriate normalization curves 
unrealistically complicated. 

\begin{figure}
\begin{center}
\includegraphics[width=0.4\hsize]{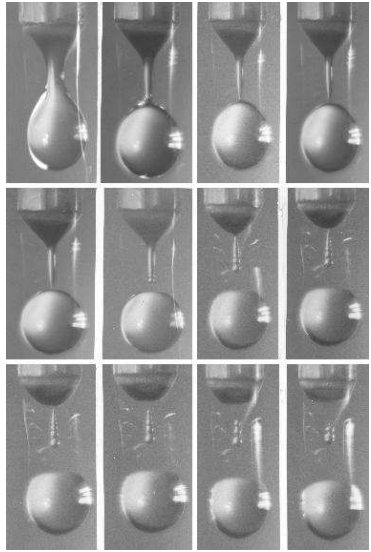}
\end{center}
\caption{\label{peregrine} 
A high-resolution sequence showing the bifurcation of a drop of 
water (\cite{PSS90}).
  }
\end{figure}

\begin{figure}
\begin{center}
\includegraphics[width=8cm]{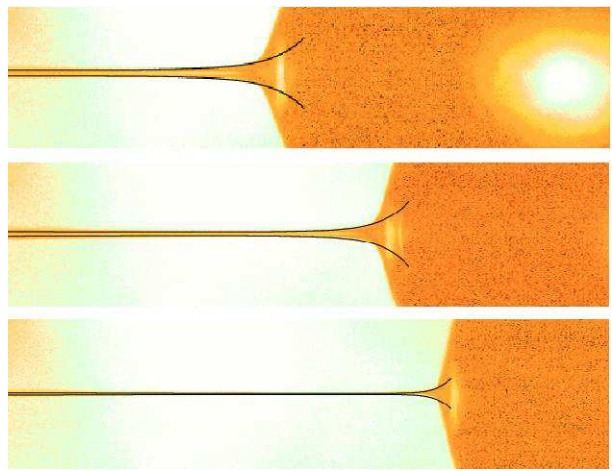}
\end{center}
\caption{\label{kowalewski} 
A sequence of interface profiles of a jet of glycerol 
close to the point of breakup (\cite{K96}). The experimental images 
correspond to $t_0 - t=350 \mu s, 298 \mu s$, and $46 \mu s$. 
Corresponding analytical solutions based on self-similarity
of the entire profile are superimposed. 
  }
\end{figure}

\section{Modern times}

After Edgerton's paper, the next paper that could report significant 
progress in illuminating {\it non-linear} aspects of drop break-up 
was published in 1990 (\cite{PSS90}). Firstly, it contains a detailed 
sequence of a drop of water falling from a pipette $D = 5.2mm$ in diameter, 
renewing efforts to understand the underlying dynamics. Secondly,
it was proposed that close to pinch-off the dynamics actually
becomes quite simple, since any {\it external} scale cannot play
a role. Namely, if the minimum neck radius $h_{min}$ is the only 
relevant length scale, and if viscosity does not enter the description,
than at a time $t_0 - t$ away from breakup on must have 
\begin{equation}
\label{inviscid}
h_{min} \propto \left(\frac{\gamma}{\rho}\right)^{2/3}(t_0 - t)^{2/3}
\end{equation}
for dimensional reasons. At some very small scale, one expects viscosity
to become important. The only length scale that can be formed from fluid
parameters alone is 
\begin{equation}
\label{ell}
\ell_{\nu} = \frac{\nu^2\rho}{\gamma}. 
\end{equation}
Thus the validity of (\ref{inviscid}) is limited to the range
$D\gg h_{min}\gg \ell_{\nu}$ between the external scale and this inner
viscous scale. 

These simple similarity ideas can in fact be extended to obtain 
the laws for the entire profile, not just the minimum radius (\cite{E93}). 
Namely, one supposes that the profile around the pinch point 
remains the same throughout, while it is only its radial and axial 
length scales which change. In accordance with (\ref{inviscid}), 
these length scales are themselves power laws in the time distance 
from the singularity. In effect, by making this transformation one
has reduced the extremely rapid dynamics close to break-up to a 
static theory, and simple analytical solutions are possible. 

The experimental pictures in Fig.\ref{kowalewski} are again taken 
using a stroboscopic technique, resulting in a time resolution
of about $10\mu s$ (\cite{K96}). Since for each of the pictures the 
temporal distance away from breakup is known, the form of the profile
can be predicted without adjustable parameters. The result of the 
theory is superimposed on the experimental pictures of a glycerol 
jet breaking up as black lines. In each picture the drop 
about to form is seen on the right, a thin thread forms on the left. 
The neighborhood of the pinch point is described quite well; in 
particular, theory reproduces the extreme {\it asymmetry} of the 
profile. We already singled out this asymmetry as responsible for 
the formation of satellite drops. 

One of the conclusions of this brief overview is that research works
in a fashion that is far from straightforward. Times of considerable 
interest in a subject are separated by relative lulls, and often known results,
published in leading journals of the day, had to be rediscovered. 
However from a broader perspective one observes a development 
from questions of (linear) stability 
and the measurement of static quantities, to a focus that is more
and more on the (non-linear) dynamics that makes fluid mechanics
so fascinating. 

\begin{acknowledgments}
I have the pleasure to acknowledge very helpful input from 
Christophe Clanet and David Qu\'er\'e.
\end{acknowledgments}


\begin{chapthebibliography}{1}
\kluwerbib

\bibitem[Savart (1833)]{S1833}
F. Savart, Ann. Chim. {\bf 53}, 337; plates in vol. 54,
(1833).

\bibitem[Clanet (2003)]{Cc}
I am relying on remarks by Christophe Clanet, a scholar
of Savart's life and achievements. 

\bibitem[Plateau (1843)]{P1843}
J. Plateau, Acad. Sci. Bruxelles M\'{e}m. {\bf XVI}, 3 (1843).

\bibitem[Plateau (1849)]{P1849}
J. Plateau, Acad. Sci. Bruxelles M\'{e}m. {\bf XXIII}, 5
(1849).

\bibitem[Hagen (1849)]{H1849}
G. Hagen, {\it Verhandlungen Preuss. Akad. 
Wissenschaften}, (Berlin), p. 281 (1849).

\bibitem[Young (1804)]{Y1804}
T. Young, Philos. Trans. R. Soc. London {\bf 95}, 65 (1804).

\bibitem[Laplace (1805)]{L1805}
P. S. de Laplace, {\em M\'{e}chanique Celeste,}
Supplement au X Libre (Courier, Paris, 1805) 

\bibitem[Rayleigh (1879)]{R1879}
Lord Rayleigh, Proc.\ London Math.\ Soc. {\bf 10}, 4 (1879).
(appeared in the volume of 1878) 

\bibitem[E\"{o}tv\"{o}s (1886)]{E1886}
L. E\"{o}tv\"{o}s, Wied. Ann. {\bf 27}, 448 (1886).

\bibitem[Quincke (1877)]{Q1877}
G.H. Quincke, Wied. Ann. {\bf 2}, 145 (1877). 

\bibitem[Lenard (1887)]{L1887}
P. Lenard, Ann. Phys. {\bf 30}, 209 (1887).

\bibitem[Rayleigh (1891)]{R1891}
Lord Rayleigh, Nature {\bf 44}, 249 (1891).

\bibitem[Bohr (1909)]{B09}
N. Bohr, Phil. Trans. Roy. Soc. A {\bf 209}, 281 (1909).

\bibitem[Harkins and Brown (1919)]{HB19}
W.D. Harkins and F. E. Brown, J. Am. Chem. Soc. {\bf 41}, 499 (1919).

\bibitem[Edgerton et al. (1937)]{EHT37}
H.E. Edgerton, E. A. Hauser, and W. B. Tucker,
J. Phys. Chem. {\bf 41}, 1029 (1937).

\bibitem[Peregrine (1990)]{PSS90}
D.H. Peregrine, G. Shoker, and A. Symon,
J.\ Fluid Mech. {\bf 212}, 25 (1990).

\bibitem[Eggers (1993)]{E93}
J. Eggers, Phys. Rev. Lett. {\bf 71}, 3458 (1993). 

\bibitem[Kowalewski (1996)]{K96}
T.A. Kowalewski, Fluid Dyn. Res. {\bf 17}, 121 (1996).

\bibitem[Rowlinson (2002)]{R02}
J.S. Rowlinson, 
{\it Cohesion}, 
Cambridge (2002). 


\end{chapthebibliography}
\end{document}